\documentclass[10pt,a4paper,fleqn]{article}
\usepackage{t1enc}
\usepackage[latin2]{inputenc}
\usepackage[english]{babel}
\usepackage{graphics}
\usepackage{multicol}
\usepackage{amsmath}
\usepackage{amssymb}
\usepackage{natbib}
\usepackage{color}

\usepackage{float}

\def\eqref#1{equation~(\ref{#1})}

\newsavebox{\boxcmdbox}

\begin{document}

\noindent
\textbf{\huge Photometry of a photometer}
\vspace*{2mm}

\noindent
{\Large Andr\'as P\'al\footnote{e-mail: apal@szofi.net}}

\noindent
{\footnotesize\it Konkoly Observatory of the Hungarian Academy of Sciences, 
        Konkoly Thege Mikl\'os \'ut 15-17,
        Budapest, H-1121, Hungary}

\noindent
{\footnotesize\it Department of Astronomy, Lor\'and E\"otv\"os University, 
        P\'azm\'any P\'eter s\'et\'any 1/A, 
        Budapest H-1117, Hungary}

\vspace*{2mm}

\begin{abstract}
\noindent
In this draft photometry and astrometry is presented from the Herschel
Space Observatory (HSO). This spacecraft orbits the second 
Lagrangian point (L2) of the Sun -- Earth system, yielding a mean distance
of a million miles ($\approx 1.5$\,million kms) for HSO. From such 
a distance, HSO is observable as a $17-23$ magnitude object moving
relatively fast (apparently several arcseconds in a minute) and the 
actual observed brightness highly depends on the spatial orientation
of the spacecraft. This draft describes briefly how
observations from this observatory and the subsequent data 
reductions have been carried out. Our conclusion is really reassuring,
namely the brightness variations of HSO are in accordance with the
publicly available reported logs and target coordinates of this spacecraft.
\end{abstract}

\vspace*{3mm}

\begin{multicols}{2}

\section{Introduction}
\label{sec:introduction}

The Herschel Space Observatory (hereafter HSO or simply Herschel) is 
a far-infrared and submillimetre observatory of the European Space
Agency (ESA), successfully launched on May 14, 2009 \citep{pilbratt2010}.
Herschel orbits the second Lagrangian point of the Sun -- Earth system,
that is apparently on the direction opposite to the Sun. The mean height
of the orbit is approximately $1.5$\,million kms, while the semi-amplitude
of the actual motion is about the half of this value. 

HSO is equipped with three on-board scientific instruments:
the Photodetector Array Camera and Spectrometer \citep[PACS,][]{poglitsch2010},
the Spectral and Photometric Imaging Receiver \citep[SPIRE,][]{griffin2010}
and the Herschel-Heterodyne Instrument for the Far-Infrared \citep[HIFI,][]{degraauw2010}.
Herschel observes a typical target from a dozen of minutes up to several
hours. The list of observed targets, the utilized instruments 
and other information can be found on the public web page of 
HSO\footnote{http://herschel.esac.esa.int/logrepgen/observationlist.do}.
Since the solar panels and other stuff mounted on the exterior of the
spacecraft reflects the sunlight efficiently, one can expect a considerably
good detectability of the spacecraft. However, due to the sharp terminations
of (some of) the external devices (see any photographs of models 
of the observatory itself), the amount of reflected light starkly 
depends on the spatial orientation of the spacecraft. Indeed, the apparent
visual brightness of the spacecraft can vary in a range of several 
magnitudes, yielding all values from the complete non-detection up to 
$R=17^{\rm m}$. In addition, as we highlight in this draft,
the observed brightness is extremely sensitive
to the orientation: even a slew of few degrees can reduce or increase
the flux coming from the spacecraft by a factor of two or three. 

Here we summarize our astrometric and photometric observations of
the spacecraft that confirms the expectations discussed above.
Our main conclusion is that the observed brightnesses significantly
correlates the orientation and these brightness variations due to 
the slew of the telescope are not ``monotonic''. The structure of 
this draft is as follows. Section~2 describes the observations, Section~3
summarizes the methods used in the data reductions while Section~4
discusses the results.

\begin{figure*}
\begin{center}
\resizebox{170mm}{!}{\includegraphics{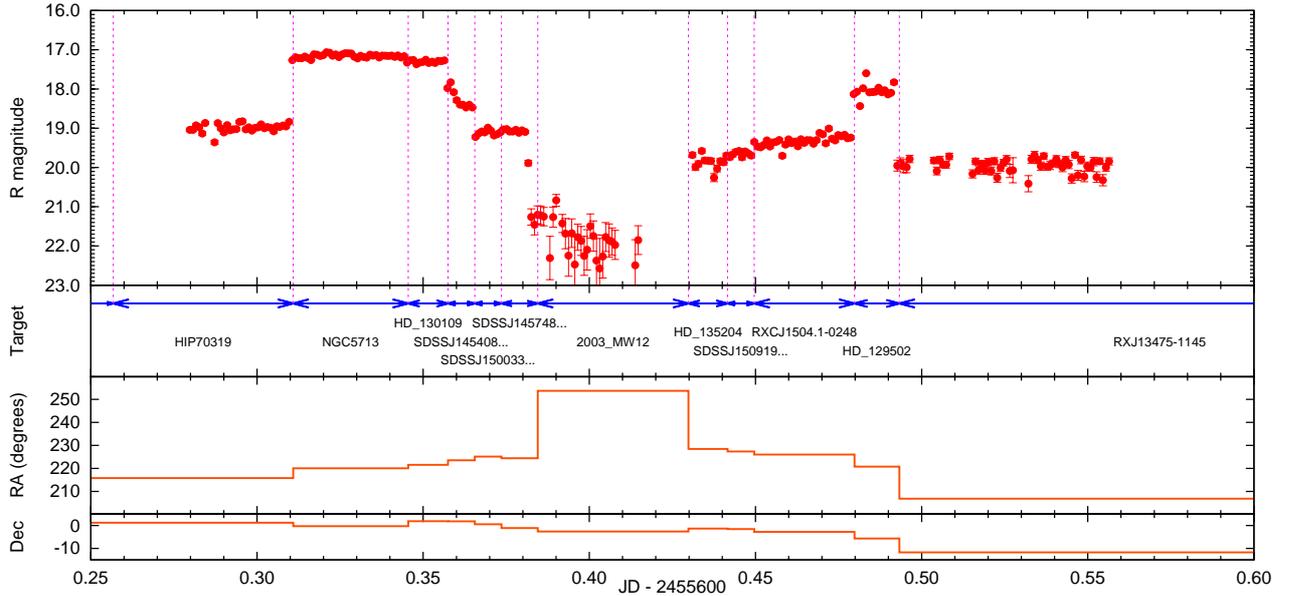}}
\end{center}\vspace*{-5mm}
\caption{The light curve of Herschel on the night of February 7/8, 2011.
The upper panel shows the light curve itself, indicating the formal
(statistical) uncertainties for each individual measurement. The 
second panel displays the then-observed targets while the two
lower panels show the celestial coordinates (right ascension and 
declination) of these targets. Note that the units for both the RA and Dec
coordinates are in degrees.} \label{fig:hsolightcurve}
\end{figure*}

\section{Observations}
\label{sec:observations}

Since apparently HSO moves relatively fast, planning of the 
observations should be done carefully. The observatory itself
is relatively faint so longer exposure times are needed for a good
astrometry and/or photometry, however, the apparent proper motion of the 
spacecraft limits the exposure times. Actually, one can compute
an ``optimal'' exposure time by comparing the instrumental FWHM with the
proper motion and let this point source move by a certain fraction
(e.g. the half or little more) of the FWHM. 

We carried out the photometric and astrometric observations of HSO using 
the the 60/90/180\,cm Schmidt telescope of the Konkoly Observatory, 
located at the Piszk\'estet\H{o} Mountain Station on the night of
February 7, 2011. The telescope is equipped with an Apogee 
ALTA-U $4{\rm k}\times 4{\rm k}$ CCD camera, yielding a square-shaped
field-of-view with a size of $1.17$\,degrees. The typical effective FWHM 
for this setup is between $2-3$ arcseconds, depending on the current
seeing. According to the predictions made available by Minor Planet Center
(MPC), the expected proper motion of Herschel was between $2.5$ and $3.5$
arcseconds per minute, therefore we concluded to use an exposure time
of 60 seconds. The observations covered the time range between 
18:43 UT and 01:21 UT (on February 8), with small gaps. All in all we 
gathered $263$ good quality scientific frames. 
In order to obtain the highest signal-to-noise
ratio, we did not use any specific filter for the observations. This
is a common practice for the astrometry (and sometimes photometry)
of small bodies in the Solar System and due to the spectral responsivity
of the employed CCD detector, such data can be interpreted
as some sort of wide R measurements. The expected total proper motion
of the object was less than 20 arcminutes, so due to the large 
field-of-view of the telescope and camera, no manual tracking 
was needed during the observations.

\begin{table*}
\caption{Samples for astrometry and photometry of Herschel,
on the night of February 7/8, 2011. This data sample below is formatted to 
the MPC standards, as it is submitted.}\label{table:sample}
\begin{center}\begin{tabular}{l}
\ttfamily	Herschel ~ ~ ~C2011 02 07.77980 07 57 15.88 +09 25 52.6 ~ ~ ~ ~ ~19.0 R ~ ~ ~561 \\
\ttfamily	Herschel ~ ~ ~C2011 02 07.78072 07 57 16.07 +09 25 49.9 ~ ~ ~ ~ ~19.0 R ~ ~ ~561 \\
\ttfamily	Herschel ~ ~ ~C2011 02 07.78163 07 57 16.26 +09 25 47.7 ~ ~ ~ ~ ~18.9 R ~ ~ ~561 \\
\ttfamily \dots
\end{tabular}\end{center}\vspace*{-5mm}
\end{table*}

\section{Data reduction}
\label{sec:datareduction}

\subsection{Calibration and initial astrometry}

Following the standard calibration procedures (subtraction of averaged
dark frames and corrections using flat field images), the reductions
of the data were done as follows. Individual stellar profiles has 
been searched using the \texttt{fistar} utility of the software package
described in \cite{pal2009}. These lists of detected stars has been used to
perform both relative and absolute astrometry on the images. The 
relative (or differential) astrometry was done with respect to a selected frame 
from the middle of the observation series (namely, the 130st out of
the 263 frames) while the absolute astrometry was done using the
USNO-B1.0 catalogue (that is also commonly used in the practice of the
astrometry of minor bodies in the Solar System). The tasks of astrometry
has been done by the \texttt{grmatch} and \texttt{grtrans} utilities. 
Both the relative and absolute astrometry were done using a third-order
polynomial fit, that has been found to be sufficient (since the 
unbiased fit residuals did not decrease if the orders were increased).

\begin{figure}[H]
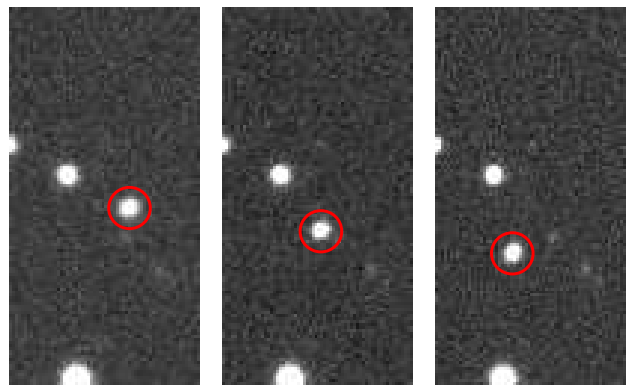

\begin{center}
\resizebox{25mm}{!}{\includegraphics{trim1.eps}}\hspace*{3mm}%
\resizebox{25mm}{!}{\includegraphics{trim2.eps}}\hspace*{3mm}%
\resizebox{25mm}{!}{\includegraphics{trim3.eps}}\hspace*{3mm}
\end{center}\vspace*{-5mm}
\caption{Herschel on its brightest phase, while observing 
NGC~5713. The image stamps cover an area of approximately $1'\times 2'$,
while the time elapse between the images is $\approx 4$\,minutes.} 
\label{fig:hsobrightest}
\end{figure}

\subsection{Registration}

The images have then been registered (using the task \texttt{fitrans})
to the reference frame using
the results yielded by the differential astrometry. 
Following the registration of the images, the flux levels of the images
have also been scaled linearly to the same level. This was done
by selecting several comparison stars and performing aperture photometry
on these (involving the task \texttt{fiphot}). 
The mean of the aperture backgrounds have been subtracted
from the respective images (yielding a zero mean background level) while
the instrumental fluxes has been scaled appropriately to the flux 
level of the reference image (that was, in practice, the same 
as the one used as an astrometric reference). A master reference image
was then built (employing the program \texttt{ficombine}),
from every tenth of the individually registered and scaled
images by taking the per-pixel median of these. 

\subsection{Astrometry and photometry of Herschel itself}

Herschel itself was then searched by eye on $7\times 3$ frames (i.e. 
3 subsequent images in 7 groups, spreading almost homogeneously 
through the whole observation). Using these coordinates as a hint
for astrometry, the task \texttt{fiphot} was used to refine 
the pixel coordinates and then a third-order polynomial has been 
fitted to both the $X$ and $Y$ coordinates as the function of the
time. We found that fitting a third order polynomial was sufficient
at a certain level (i.e. the residuals were not larger than few tenths
of a pixel) and this polynomial was used to interpolate the current position
of the HSO between the frames were it was marginally or absolutely not
detected. This interpolated coordinates have been exploited as a hint
for a subsequent centroid determination (using the task \texttt{fiphot}).
These refined centroid coordinates were then used as a photometric
centroid, on which the actual aperture photometry was performed. 
In order to avoid the effect of nearby stars (that were crossed by
Herschel apparently), this process has been done on differential
images (i.e. the master reference image has been subtracted from
each registered and scaled image before astrometry and photometry).

\subsection{Almost-standard magnitudes}

To have an accurate photometry, all of the nicely detected stars on the
master reference image has been measured similarly (with the
same aperture size as for Herschel) and using 
the cross-match lists from the absolute astrometry, we were able
do determine the shift between the standard (USNO) and the instrumental
magnitudes. Although this step sounds a bit optimistic (i.e. the flux
levels derived from a CCD image without any filter has been adjusted
to a photographic reference catalogue with a given filter and we 
neglect the effects of the color-dependent extinction),
we found that the residual of this fit is not larger than $0.07^{\rm m}$.
On can see the resulted light curve on Fig.~\ref{fig:hsolightcurve}.
Image stamps showing Herschel itself on some of its brightest phases
can be seen on Fig.~\ref{fig:hsobrightest}, while the results of
astrometry and photometry in the format of MPC are displayed on
Table~\ref{table:sample}. 

\section{Discussion}
\label{sec:discussion}

In this draft some photometric and astrometric measurements for the 
Herschel Space Observatory have been presented. As it was discussed here,
taking accurate photometry of HSO has some difficulties due to the fast 
apparent proper motion. Anyway, the presented photometric results
confirm that the brightness variations of HSO (yielded by the telescope slews)
is in accordance with the observation logs, and\dots and this is really reassuring.

\section*{Acknowledgements}
\label{sec:acknowledgements}

The author would thank J\'anos Kelemen for the discussions and gathering 
the observations and Kriszti\'an S\'arneczky for useful advises. 
The scientific work of the author related to the Herschel mission
has been supported by the ESA grant PECS~98073 and in part by 
the J\'anos Bolyai Research Scholarship of the Hungarian Academy of Sciences.

{}
\end{multicols}

\end{document}